\documentclass[12pt]{article}
\pdfoutput=1
\usepackage{amsmath, amssymb,amsthm,amscd,graphicx,dsfont,amsthm,youngtab}
\usepackage{mathrsfs}
\usepackage{hyperref}
\usepackage{subfigure}
\usepackage{mathtools}
\usepackage{tabu}
\usepackage{bigj}
\usepackage{tikz-cd}

\usepackage{tikz}
\usetikzlibrary{arrows,decorations.pathmorphing,decorations.pathreplacing,decorations.markings,shapes.geometric,calc,positioning,chains,matrix,scopes} 
\numberwithin{equation}{section}% [if desired]
\theoremstyle{definition}
\newtheorem{thm}{Theorem}[section]

\theoremstyle{definition}

\theoremstyle{definition}
\newtheorem*{conj*}{Conjecture}

\theoremstyle{remark}

\newcommand{\R}{\mathbb{R}}

\newcommand{\bC}{\mathbb{C}}

\renewcommand{\P}{\mathbb{P}}
\newcommand{\Z}{\mathbb{Z}}

\newcommand{\g}{\mathfrak{g}}

\renewcommand{\L}{\mathcal{L}}
\renewcommand{\O}{\mathscr{O}}
\newcommand{\E}{\mathcal{E}}

\newcommand{\cT}{\mathcal{T}}
\newcommand{\M}{\mathcal{M}}

\newcommand{\Phibar}{\overline{\Phi}}
\newcommand{\Dbar}{\overline{D}}

\newcommand{\pp}[2]{\frac{\partial #1}{\partial #2}}

\DeclareMathOperator{\Tr}{Tr}

\DeclareMathOperator{\Hom}{Hom}

\DeclareMathOperator{\Gr}{Gr}
\DeclareMathOperator{\Pf}{Pf}

\DeclareMathOperator{\Exp}{Exp}

\DeclareMathOperator{\Sym}{\mathrm{Sym}}

\begin{document}

\begin{titlepage}

\begin{center}
{\Large \bfseries
Local Operators from the Space of Vacua of Four Dimensional SUSY Gauge Theories
}

\vskip 1.2cm
Richard Eager \footnote{eager@mathi.uni-heidelberg.de}
\bigskip
\bigskip

%\begin{tabular}{ll}
Mathematical Institute, \it Heidelberg University, Heidelberg, Germany \\[.1cm]
%\end{tabular}

\vskip 1.5cm

\textbf{Abstract}
\end{center}

\medskip
\noindent
We construct local operators in short representations of supersymmetry algebras from polyvector fields on the quantum moduli space of vacua of supersymmetric gauge theories.  These operators form a super Lie algebra under a natural bracket operation with structure constants determined by terms in the operator product expansion of the corresponding operators.  We propose a formula for the superconformal index in terms of an index over polyvector fields on the moduli space of vacua.  

Along the way, we construct several models with moduli space of vacua corresponding to affine cones over smooth bases using the classical geometry of Severi varieties
and the Landsberg-Manivel projective geometries corresponding to the Freudenthal magic square of exceptional Lie algebras.  Curiously, we relate the Landsberg-Manivel projective geometries to the exceptional enhanced symmetry surprises of Dimofte and Gaiotto.  Finally, we determine Beasley-Witten higher-derivative F-terms in new examples arising from Severi varieties and remark on their origin in classical projective duality.

\bigskip
\vfill
\end{titlepage}

\tableofcontents

\newpage 
%----------------------------------------------------------
\section{Introduction}
%----------------------------------------------------------
Supersymmetric quantum field theories have furnished a rich testing ground for many ideas in quantum field theory.  The theories often have classical flat directions, and as a result, the theories often have a space of inequivalent vacua called the classical moduli space.  Remarkably, the strong constraints of supersymmetry often allow the quantum corrected moduli space of vacua to be exactly determined \cite{Seiberg:1994bz}.    Supersymmetric quantum field theories are often connected by a rich web of electric-magnetic Seiberg dualities that relate strongly coupled gauge dynamics in one theory to weakly coupled gauge dynamics in the dual theory \cite{Seiberg:1994pq}.  Traditionally, these dualities have been studied by `t Hooft anomaly matching, comparing deformations, matching local operators, and matching the quantum moduli space of vacua \cite{Seiberg:1994pq}.  

In this paper, we attempt to derive the spectrum of BPS local operators directly from the quantum moduli space of vacua.  In the rare instances when the moduli space of vacua is an affine complex cone over a smooth K\"ahler base, we are surprisingly successful.  We determine the operators that are BPS with respect to a fixed supercharge.  As a result, we can recover the superconformal index directly from the quantum moduli space of vacua.  Our basic strategy is to view the low energy effective theory as an $\mathcal{N}=1$ supersymmetric nonlinear sigma model from Minkowski space or $\R \times S^3$ to $\mathcal{M},$ the quantum moduli space of vacua.  The local operators arise as cohomology classes of polyvector fields on $\mathcal{M}$, which is familiar as the ring of local observables in the topological B-model on $\mathcal{M}.$  These operators had largely been ignored in supersymmetric QCD (SQCD) until Beasley and Witten's study of multi-fermion F-terms \cite{Beasley:2004ys}.  We systematically compute these operators using an extension of the Borel-Weil-Bott theorem.

Using the spectrum of BPS local operators, we find a candidate expression for the superconformal index $\mathcal{I}(t,y)$ in terms of the moduli space of vacua as
$$(1-t y)(1-t y^{-1}) \Exp^{-1}[ \mathcal{I}(t,y)]  = \Exp^{-1}[\chi(t)(1 - t^2) + t^2] + \dots,$$
where $\chi(t)$ is an alternating sum of Euler characteristics of polyvector fields on the moduli space of vacua that will be defined more precisely and $\Exp^{-1}$ is the inverse of the plethystic exponential.  The formula is typically not exact due to extra degrees of freedom arising from the singularity at the origin of the moduli space \cite{Seiberg:1994bz}.  However, in the examples we consider, the formula agrees remarkably well with the superconformal index.

A similar expression for the superconformal index was found for large-N gauge theories dual to type IIB string theory on $AdS_5 \times L^5$ with $L$ Sasaki-Einstein in \cite{Eager:2012hx}.  There, the local operators were studied using cyclic homology.  The local operators in the large-N quiver gauge theory are related to polyvector fields on the Calabi-Yau cone $Y$ over $L$ via the Hochschild-Kostant-Rosenberg theorem \cite{MR0142598}.  The moduli space of vacua in these theories is roughly the large $N$ limit of $\Sym^N Y$, which is $N$-th symmetric product of $Y$.  A similar relation was also found in \cite{Costello:2016mgj, Eager:2018oww}.

Alternatively, we can determine the local operators from gauge theory using the cohomology of a nilpotent supercharge $Q.$  The classical Q-cohomology can be reformulated in terms of a generalization of Lie algebra cohomology.  This classical space of states coincides with the local operators in the holomorphic twist of the theory \cite{Costello:2011np, Costello:2015sma}.  However, the differential of the quantum corrected Q-cohomology is different due to the Konishi anomaly.  After taking into account the Konishi anomaly \cite{Konishi:1983hf}, we find that the leading contributions to Q-cohomology match the space of local operators computed from polyvector fields on the quantum moduli space of vacua in accordance with \cite{Nakayama:2006ur}.

The quantum behaviour of SQCD was studied long ago using instanton techniques.  For $N_F = N_C - 1$, instantons generate a superpotential and deform the classical moduli space of vacua.  The Konishi anomaly provides a consistency check on these calculations and can partially simplify them.  It is therefore reassuring that the Konishi anomaly also corrects the gauge theory Q-cohomology in order to match the polyvector fields on the quantum moduli space of vacua.  Similarly, the multi-fermion operators studied by Beasley and Witten arise from instanton effects in SQCD.

Finally, we verify that our counting of operators is consistent with the superconformal index.  The superconformal index \cite{Romelsberger:2005eg, Kinney:2005ej,Romelsberger:2007ec} has been used to match protected operators in dual quantum field theories.  In a superconformal theory, the index counts protected operators satisfying a BPS condition that cannot be combined to form long multiplets.  The equality of the index for a theory and its electric-magnetic dual often lead to very interesting integral identities between products of elliptic gamma functions \cite{Dolan:2008qi, Spiridonov:2008zr}.  Conversely, recently discovered integral identities have led to the derivation of new dualities between quantum field theories.  One of the great features of the superconformal index is that it can be easily computed, at least in a perturbative expansion.  Part of its simplicity follows from its construction, which is insensitive to the precise form of the superpotential.  However, the Q-cohomology does depend on the explicit form of the superpotential.  

One of the main motivations of this study is to develop a strategy to prove that the $Q$-cohomology groups of local operators in two Seiberg dual theories are isomorphic.  This can be viewed as a categorification of the equality of superconformal indices and in particular Spiridonov's elliptic beta integral \cite{MR1846786}.  
Since the superconformal index is the partition function for the holomorphically twisted theory \cite{Eager:2018oww}, it is natural to conjecture that even more is true.  Namely, not just the operators, but the correlation functions should be equivalent.  In the language of holomorphic factorization algebras developed by Costello and Gwilliam \cite{MR3586504}, our conjecture is that 
\begin{conj*}
The holomorphic factorization algebras associated to the holomorphic twists of two Seiberg dual theories are (quasi-)isomorphic.
\end{conj*}
Since two Seiberg dual theories have equivalent quantum moduli spaces of vacua, it is natural to prove the equality of the $Q$-cohomology groups by relating both cohomologies to the local operators that can be described directly on the quantum moduli space of vacua.
In two-dimensions, Ando and Sharpe \cite{Ando:2009av} showed that the superconformal index (elliptic genus) of a Landau-Ginzburg model is equal to that of its low-energy sigma model  
using a Thom class computation.  It is natural to expect that a similar result holds in four dimensions. 
The full superconformal index can be derived using the holomorphically twisted sigma model appearing in \cite{Aganagic:2017tvx}.  The calculations in this paper represent the first steps toward evaluating these indices.  We find many indications that there might be simplifications in the full formula, perhaps arising from cohomology vanishing theorems.

Identifying local operators in terms of polyvector fields has the added virtue that we can adapt several classical results on polyvector fields to gauge theory.  In particular, polyvector fields have a Schouten-Nijenhuis bracket operation that generalizes the ordinary Lie bracket of vector fields.  On a complex manifold, the Schouten-Nijenhuis bracket on global sections of the sheaf of polyvector fields extends to a Gerstenhaber algebra on the cohomology of the sheaf of polyvector fields.  Following a suggestion of Costello, we propose that the Shouten-Nijenhuis bracket computes certain protected OPE coefficients.  These can be viewed as the leading OPE coefficients in the holomorphically twisted theory \cite{Costello:2011np, Costello15}.  The bracket is an example of a secondary products in supersymmetric field theory \cite{Beem:2018fng}. 

Many of the techniques developed here can be applied to theories in various dimensions and varying amounts of supersymmetry.  In particular in two dimensions, a gauge theory that has a Calabi Yau manifold $X$ as its moduli space of vacua in the IR, has the same elliptic genus as the sigma model with target $X.$  The elliptic genus is the two dimensional analog of the superconformal index, and it is explicitly expressible in terms of vector bundles on the Calabi Yau target \cite{Witten:1986bf, MR970288}.

%----------------------------------------------------------
\section{Overview}
%----------------------------------------------------------
Typically, supersymmetric theories have moduli spaces of vacua with several branches and high-dimensional singular loci.  To simplify the analysis of the physics at the singular loci, we first consider models where the moduli space of vacua is an affine complex cone over a smooth K\"ahler base.  In this case, the only singularity is at the origin of the cone.  These models are not very common.  We consider two types of families of theories with moduli space of vacua that are affine cones of a smooth K\"ahler base.  The first family is SQCD with gauge group $SU(2)$ and $N_f$ flavors.  Its moduli space of vacua is an affine cone over the Grassmannians $\Gr(2, 2 N_f).$  The second family consists of affine cones over the four Severi varieties and their hyperplane sections.  They can be described by generalized Wess-Zumino models with cubic superpotentials.  Surprisingly all of these models arise as projective geometries associated to Freudenthal's magic square \cite{MR1832903}.  The most complicated member of the family is the Cayley plane $\mathbb{OP}^2,$ which was recently considered in \cite{Razamat:2016gzx}.

Finally, we will consider models with singular loci.  Our main tool will be to utilize the geometry of orbit closures.  While seemingly esoteric, the geometry of orbit closures elegantly recovers the structure of the moduli space of $Sp(n)$ SQCD.  For exceptional Lie algebras, the geometry of orbit closures will be used to describe the various smooth and singular strata of moduli spaces given by superpotentials of degree four or more.

When the quantum moduli space of vacua $\M$ is an affine cone over a smooth projective K\"ahler base $B,$ we review how the polyvector fields on $B$ can be pulled back to polyvector fields on $\M$ following \cite{Beasley:2004ys}.  We will construct local operators from cohomology classes $H^{\bullet}(B, \wedge^{\bullet} T B \otimes \mathcal{L}^k)$ where $\mathcal{L}$ is the line bundle obtained from pulling back $\mathcal{O}(1)$ to $B$.

We find that the alternating sum of the Euler characters of polyvector fields on $\M$ have a particularly simple expression when $\M$ is a cone over a Severi variety.  This includes both the Grassmannian $\Gr(2,6)$ and the Cayley plane $\mathbb{OP}^2$.  The simple expression arises from a small correction due to removing the identity operator and taking the plethystic logarithm.  From this index, we can determine the superconformal index and conjecture a general form of the index in terms of polyvector fields.  We then test this conjecture for SQCD with gauge group $SU(2)$ and four flavors.  We find an $E_7$ surprise where the plethystic logarithm is almost entirely expressible of characters of the exceptional Lie group $E_7$ \cite{Dimofte:2012pd}.  While the $E_7$ symmetry has a natural explanation in terms of a five dimensional gauge theory or coupling two four dimensional gauge theories together \cite{Dimofte:2012pd}, it would be desirable to have an explanation entirely in terms of the geometry of the Grassmannian $Gr(2,8).$ 
%----------------------------------------------------------
\section{Smooth models}
%----------------------------------------------------------

%----------------------------------------------------------
\paragraph{Freudenthal's magic square}
%----------------------------------------------------------
From a pair $(\mathbb{A}, \mathbb{B})$ of normed real division algebras $\mathbb{A}, \mathbb{B} \subseteq \{\R, \bC, \mathbb{H}, \mathbb{O} \}$ Freudenthal and Tits define a Lie algebra
$$\g(\mathbb{A}, \mathbb{B}) = Der \mathbb{A} \oplus (\mathbb{A}_0 \otimes \mathcal{J}_3(\mathbb{B})_0) \oplus Der \mathcal{J}_3(\mathbb{B}),$$
where $\mathbb{A}_0$ is the space of imaginary elements of $\mathbb{A}$, $\mathcal{J}_3(\mathbb{B})$ is the Jordan algebra of $3 \times 3$ $\mathbb{B}$-Hermitian matrices, and $\mathcal{J}_3(\mathbb{B})_0$ is the subspace of traceless matrices.  The corresponding Lie algebras are shown in table \ref{tab:MagicSquare}.

\begin{table}[htp]
\begin{center}
\begin{tabular}{c|cccc}
 & $\R$ & $\bC$ & $\mathbb{H}$ &  $\mathbb{O}$ \\
\hline
$\R$ & $\mathfrak{so}_3$ & $\mathfrak{sl}_3$ & $\mathfrak{sp}_6$ & $\mathfrak{f}_4$ \\
$\bC$ & $\mathfrak{sl}_3$ & $\mathfrak{sl}_3 \times \mathfrak{sl}_3$ & $\mathfrak{sl}_6$ & $\mathfrak{e}_6$ \\
$\mathbb{H}$ & $\mathfrak{sp}_6$ & $\mathfrak{sl}_6$ & $\mathfrak{so}_{12}$ & $\mathfrak{e}_7$ \\
$\mathbb{O}$ & $\mathfrak{f}_4$ & $\mathfrak{e}_6$ & $\mathfrak{e}_7$ & $\mathfrak{e}_8$ \\
\end{tabular}
\end{center}
\caption{Freudenthal's Magic Square}
\label{tab:MagicSquare}
\end{table}%
To each Lie algebra in Freudenthal's magic square, Landsberg and Manivel \cite{MR1832903} associate a projective geometry.  These geometries are listed in table \ref{tab:MagicGeom}.  The Lie algebra of the projective geometry's isometry group is the corresponding Lie algebra in the magic square.
\begin{table}[htp]
\begin{center}
\begin{tabular}{c|cccc|c}
 & $\R$ & $\bC$ & $\mathbb{H}$ &  $\mathbb{O}$ & Family Name\\
\hline
$\R$ & $\nu_2(Q^1)$ & $\P(T \P^2)$ & $Gr_{\omega}(2,6)$ & $\mathbb{OP}_0^2$ & hyperplane section of Severi \\
$\bC$ & $\nu_2(\P^2)$ & $\P^2 \times \P^2$ & $Gr(2,6)$ & $\mathbb{OP}^2$ & Severi variety\\
$\mathbb{H}$ & $Gr_{\omega}(3,6)$ & $Gr(2,6)$ & $\mathbb{S}_{12}$ & $E_7/P_7$ & lines through a points of $G^{ad}$ \\
$\mathbb{O}$ & $F_4^{ad}$ & $E_6^{ad}$ & $E_7^{ad}$ & $E_8^{ad}$ &  $G^{ad}$ \\
\end{tabular}
\end{center}
\caption{Freudenthal Geometries}
\label{tab:MagicGeom}
\end{table}%
All of the ``exceptionally simple exceptional models'' studied by Razamat and Zafrir \cite{Razamat:2016gzx} can be naturally associated to one of the Manivel-Landsberg projective geometries.  The first two families of Severi varieties and their hyperplane sections are described by the critical locus of a cubic superpotential.
The third family, also known as the ``subexceptional series,'' describes varieties that are the critical locus of a quartic superpotential.  The fourth and final row corresponds to the Deligne-Cvitanovi\'c exceptional series and appears in the classification of rank one $\mathcal{N} = 2$ SCFTs \cite{Beem:2017ooy}.  

The third family of Freudenthal Lie groups appears in another interesting way.
Dimofte and Gaiotto found an $E_7$ ``surprise'' in the theory of 28 five-dimensional free hypermultiplets coupled to four-dimensional SQCD with 
four flavors (eight chiral multiplets).  The flavor symmetry group enhanced from $SU(8)$ to $E_7.$  Similarly, they found $SO(12)$ enhanced symmetry with 16 four-dimensional free hypermultiplets coupled to three-dimensional SQCD with three flavors.  These two enhanced symmetry groups appear in the third row of the Freudenthal magic square.  It is natural to conjecture that there is an enhanced $SU(6)$ flavor symmetry group for 10 three dimensional matter multiplets coupled to two dimensional SQCD with two flavors extending the pattern.
%----------------------------------------------------------
\paragraph{Severi varieties}
%----------------------------------------------------------
Affine cones over the four Severi varieties arise as the moduli space of vacua of simple generalized Wess-Zumino models consisting of $n$ chiral multiplets with a specific cubic superpotential.  The four Severi varieties correspond to the four real division algebras $\mathbb{A} = \R, \bC, \mathbb{H},$ and $\mathbb{O}.$  The cubic polynomial arises as the determinant of symmetric three-by-three matrix over the corresponding division algebra.  It is therefore a cubic polynomial in $3 + 3 \dim \mathbb{A}$ real variables, which is the number of chiral multiplets.  Since the superpotential has $r$-charge 2, the chiral multiplets have $r$-charge 2/3.  The models have an extended flavor symmetry group $G,$ and the chiral multiplets are in a representation $V_{\lambda}$ of $G$.  The corresponding groups are listed in Table \ref{tab:Severi}.    The algebras $\R$ and $\bC$ correspond to the Veronese embedding of $\P^2$ in $\P^5$ and the Segre embedding of $\P^2 \times \P^2$ in $\P^8$ respectively.  The algebra $\mathbb{H}$ corresponds to the Grassmannian $\Gr(2,6)$.  Finally the algebra $\mathbb{O}$ corresponds to the Cayley plane.
\begin{table}[htp]
\begin{center}
\begin{tabular}{|c|c|c|c|c|c|}
\hline
\# chirals& $X$ & $\dim(X)$ & $G_{Flavor}$ & Matter Rep $V_{\lambda}$ \\
\hline
6 & $\P^2 \subset \P^5$ & 2 & $SU(3)$ & $\Sym^2 \mathbf{3}$ \\
\hline
9 & $\P^2 \times \P^2 \subset \P^8$ & 4 & $SU(3) \times SU(3)$ & $\mathbf{3} \otimes \overline{\mathbf{3}}$ \\
\hline
15 & $\Gr(2,6)$ & 8 & $SU(6)$ & $\wedge^2 \mathbf{6}$ \\
\hline
27 & $\mathbb{OP}^2$ & 16 & $E_6$ & $\mathbf{27}$ \\
\hline
\end{tabular}
\end{center}
\caption{Wess-Zumino models of Severi Varieties}
\label{tab:Severi}
\end{table}%
%----------------------------------------------------------

The four Severi varieties arise as the projective duals of cubic hypersurfaces $$X^{\vee (\dim(V_{\lambda}))}_3 \in \P^{\dim(V_{\lambda}) - 1}$$
in $\dim(V_{\lambda}) - 1$ dimensional projective space \cite{MR2113135}.
The cubic polynomial defining the hypersurface is the cubic superpotential.  The space of cubics polynomials has dimension
$${\dim(V_{\lambda}) + 2 \choose 3} - \dim(V_{\lambda})^2.$$  This is equal to the dimension of the conformal manifold of the corresponding Wess-Zumino theory \cite{Green:2010da}.

%----------------------------------------------------------
\subparagraph{Grassmannian $Gr(2,6)$}
%----------------------------------------------------------
The most familiar member of the Severi varieties is the Grassmannian $Gr(2,6)$.  It arises as the moduli space of vacua of $SU(2)$ QCD with three flavors.
$SU(2)$ QCD with three flavors has a magnetic dual consisting of 15 chiral multiplets $M^{ij}$ and cubic superpotential
$$W = \epsilon_{i_1 j_1 i_2  j_2  i_3 j_3} M^{i_1 j_1} M^{i_2 j_2} M^{i_3 j_3}.$$
The F-term relations are the Pl\"ucker relations for the Grassmannian $Gr(2,6)$.
The moduli space of vacua is
$$\M = \bC[M^{ij}]/ \left(\epsilon_{i_1 j_1 i_2  j_2  i_3 j_3} M^{i_1 j_1} M^{i_2 j_2}  \right).$$
This moduli space is an affine cone over the Grassmannian $Gr(2,6)$.
Finally, there is one syzygy that arises since all of the relations are obtained from a superpotential.
In general, the Pl\"ucker embedding for $Gr(n,2)$ has $n \choose 2$ Pl\"ucker variables, $n \choose 4$ Pl\"ucker relations,
and $n \choose 6$ syzygies.  Only for $Gr(2,6)$ do the number of  Pl\"ucker variables and Pl\"ucker relations agree as required for a theory with only chiral multiplets.  It is then an extra condition that there is only one syzygy and no higher syzygies.

%----------------------------------------------------------
\subparagraph{Cayley plane $\mathbb{OP}^2$}
%----------------------------------------------------------
A similar model to the Grassmannian $Gr(2,6)$ was considered in \cite{Higashijima:1999ki} and more recently in \cite{Razamat:2016gzx}.  It consists of 27 chiral multiplets $\Phi$ with an $E_6$ invariant superpotential $W = W_{E_6}(\Phi).$  The cubic polynomial $W_{E_6}$ was first written down by Cartan in his thesis \cite{CatanThesis}.  Another more recent description is in \cite{MR1854697}.
The moduli space of vacua is an affine cone over the Cayley plane $\mathbb{OP}^2$.
This implies that the Cayley plane is also described by 27 variables, 27 relations, and one syzygy.
The relations are described in \cite{Lax} and arise from the appearance of the $\overline{\mathbf{27}}$ in $\wedge^2 \mathbf{27} \cong \mathbf{351} \oplus \overline{\mathbf{27}}$.
%----------------------------------------------------------
\paragraph{Hyperplane sections of Severi varieties}
%----------------------------------------------------------
The hyperplane sections of Severi varieties also occur as the critical locus of a cubic superpotential.  Their geometry is described in \cite{MR1966752} Section 6.3.
The flavor symmetry groups are obtained by ``folding'' the corresponding Dynkin diagram in the series of Severi varieties.  The number of chiral multiplets, the dimension of the moduli space of vacua, the corresponding flavor symmetry group, and matter representation of the flavor symmetry group are displayed in Table \ref{tab:hyperplanesection}.
\begin{table}[htp]
\begin{center}
\begin{tabular}{|c|c|c|c|c|c|}
\hline
\# chirals& $X$ & $\dim(X)$ & $G_{Flavor}$ & Matter Rep $V_{\lambda}$ \\
\hline
5 & $\nu_2(Q^1)$ & 1 & $SO(3)$ & $\Sym^2_0 \mathbf{3}$ \\
\hline
8 & $\P(T \P^2)$ & 3 & $SU(3)$ & $\mathbf{adj}$ \\
\hline
15 & $\Gr_{\omega}(2,6)$ & 7 & $Sp(6)$ & $\wedge^2 \mathbf{6}$ \\
\hline
26 & $\mathbb{OP}^2_0 = F_4/P_4$ & 15 & $F_4$ & $\mathbf{26}$ \\
\hline
\end{tabular}
\end{center}
\caption{Hyperplane sections of Severi Varieties}
\label{tab:hyperplanesection}
\end{table}%

%----------------------------------------------------------
\section{Geometry of orbit closures}
%----------------------------------------------------------
Given a group $G$ and a representation $V,$ the action of the group $G$ provides a decomposition of $V$ into orbits.  If view $G$ as a global symmetry group and $V$ as a matter representation, then the orbits correspond to various symmetry breaking patterns.  In general, the geometry of these orbits is quite complicated.  Classical invariant theory provides invariants that can distinguish between the various orbits.  The invariants can be thought of as Landau-Ginzburg order parameters.  The irreducible representations of (reductive) groups with finitely many orbits were classified by Kac in \cite{MR575790}.  Quite remarkably, almost all of these representations arise from gradings on Lie algebras.  One common way the gradings arise is from a Dynkin diagram with a distinguished node.

Our interest is that orbit closures often arise as moduli spaces of vacua in supersymmetric gauge theory.  When the largest orbit is a hypersurface defined by a polynomial $W = 0,$ the moduli space of vacua for the theory of free chiral multiplets with superpotential $W$ is a smaller orbit closure.  The results of \cite{Kras12, MR1988690} describe the geometry of the moduli space of vacua and allow us to determine its Hilbert series.

If we gauge $G$, then many of the theories we study have played an important role in the dynamics of $\mathcal{N} = 1$ supersymmetric gauge theory.  Many of these connections are described in \cite{Dotti:1997wn}.  We first illustrate how orbit closures describe the strata of the moduli space of vacua in $Sp(n)$ SQCD and then describe more exotic orbit closures.
%----------------------------------------------------------
\paragraph{$Sp(n)$ SQCD}
%----------------------------------------------------------
Let $W$ be a vector space of dimension $m$ and let $M = \Lambda^2 W$ be the space of bivectors.  The group $GL(W)$ acts on the projective space
$\P(\Lambda^2 W)$.  The nontrivial orbits range in dimension from 2 to $2 \lfloor m/2 \rfloor.$  The {\it k-th} Pfaffian variety $\Pf(2 k, W)$ is the orbit closure of bivectors with rank $2 k.$  The Pfaffian variety $\Pf(2, W)$ is the Grassmannian $\Gr(2,W)$ of $2$-planes in $W.$  The orbits are nested as
$$0 \hookrightarrow \Pf(2, W) \hookrightarrow \Pf(4,W) \dotsm \hookrightarrow \Pf(2 \lfloor m/2 \rfloor - 2, W) \hookrightarrow \Pf(2 \lfloor m/2 \rfloor, W) = \P(\Lambda^2 W).$$
In $Sp(N_C)$ SQCD with $k = N_f > N_c$ flavors, the moduli space of vacua is the Pfaffian variety $\Pf(2N_c, 2N_f)$ \cite{Intriligator:1995ne}.

For $Sp(2)$ with $N_f = 4$, the moduli space of vacua is the Pfaffian variety $\Pf(4,8).$  However, this variety occurs as the singular locus of $\Pf(6,8).$
Since $\Pf(6,8)$ is described by the vanishing of a single Pfaffian, the singular locus is given by the simultaneous vanishing of the partial derivatives of a single quartic polynomial.  This is precisely the ``magnetic dual'' description.

Despite the fact that the theory is not superconformal, we can formally consider the superconformal index given by 28 free chiral multiplets with $R$-charge $1/2.$
Then the index is
$$\mathcal{I}_{Sp(2), N_f = 4}(t,y) = 1 + 28t^{1/2} + 406 t + (4032 +28 \chi_2(y))t^{3/2} + (30681 + 784 \chi_2(y)) t^2 + \mathcal{O}(t^5/2)$$
The leading terms come from the Hilbert series of the Pfaffian variety \cite{MR2037100}
$$\frac{1+6 t^{1/2}+21 t+28 t^{3/2}+21 t^2+6 t^{5/2}+t^3}{(1-t^{1/2})^{22}} = 1+28 t^{1/2}+406 t+4032 t^{3/2}+30744 t^2 + \mathcal{O}(t^5/2)$$
and the coefficient of the $t^2$ term is $30744 - 63$, where $dim(SU(2N_f)) = 63$ is the flavor symmetry group.
%----------------------------------------------------------
\paragraph{Subexceptional series}
%----------------------------------------------------------
There is a family of quartic superpotentials corresponding to the subexceptional series of Lie groups.
\begin{table}[htp]
\begin{center}
\begin{tabular}{|c|c|c|c|c|c|}
\hline
\# chirals& $X$ & $\dim(X)$ & $G_{Flavor}$ & Matter Rep $V_{\lambda}$ \\
\hline
14 & $\Gr_{\omega}(3,6)$ & 6 & $Sp(6)$ & $ \mathbf{14}$ \\
\hline
20 & $\Gr(3,6)$ & 9 & $SU(6)$ & $\wedge^3 \mathbf{6}$  \\
\hline
32 & $\mathbb{S}_{12}$ & 16 & $SO(12)$ & $\mathbf{32}$ \\
\hline
56 & $E_7/P_7$ & 28 & $E_7$ & $\mathbf{56}$ \\
\hline
\end{tabular}
\end{center}
\caption{Subexceptional Series}
\label{tab:pd}
\end{table}%
Analogous to $Sp(n)$ SQCD, these theories are sigma models on an orbit closure.
For the theory with 20 chiral multiplets, the orbit structure is given by the following inclusion of strata \cite{Kras12}
\begin{equation}
\begin{tikzcd}
\text{Dimension} & 0 & 10 & 15 & 19 & 20 \\
\text{Orbit}          & \mathcal{O}_0 \arrow[r, dash] & \mathcal{O}_1 \arrow[r, dash] & \mathcal{O}_2 \arrow[r, dash] & \mathcal{O}_3 \arrow[r, dash] & \mathcal{O}_4 \\
\end{tikzcd}
\end{equation}
The orbit closure $\overline{\mathcal{O}}_3$ is a quartic hypersurface in $\P^{20}.$  Letting the quartic polynomial be the superpotential, the moduli space of vacua is the orbit closure $\overline{\mathcal{O}}_2$.
For the theory with 56 chiral multiplets, the orbit closures are \cite{Benedetti:2018}
\begin{equation}
\begin{tikzcd}
\text{Dimension} & 0 & 28 & 45 & 55 & 56 \\
\text{Orbit}          & \mathcal{O}_0 \arrow[r, dash] & \mathcal{O}_1 \arrow[r, dash] & \mathcal{O}_2 \arrow[r, dash] & \mathcal{O}_3 \arrow[r, dash] & \mathcal{O}_4 \\
\end{tikzcd}
\end{equation}
The orbit closure $\overline{\mathcal{O}}_3$ is a quartic hypersurface in $\P^{56}.$  Again using this quartic polynomial as a superpotential for 56 chiral multiplets with $R$-charge 1/2, the moduli space of vacua is the orbit closure $\overline{\mathcal{O}}_2$.  The orbit closure $\overline{\mathcal{O}}_1$ is the Freudenthal variety $E_7/P_7.$  The moduli space of vacua was incorrectly identified as the Freudenthal variety in \cite{Higashijima:1999ki}.
Using the free resolution in \cite{Benedetti:2018}, the Hilbert series of $\overline{\mathcal{O}}_2$ is straightforward to compute.  Combining this with the isometry group $E_7$ of the orbit closure, we can match the leading terms of the index.

\section{$SU(2)$ SQCD review}
Supersymmetric QCD with gauge group $SU(N_c)$ and $N_f$ massless flavors of quarks has a $\mathcal{N} = 1$ vector multiplet and $N_f$ chiral multiplets transforming in the fundamental representation of $SU(N_c)$ and $N_f$ chiral multiplets transforming in the anti-fundamental representation of $SU(N_c).$  The theory has a $SU(N_C) \times SU(N_C) \times U(1)_B \times U(1)_R$ global symmetry.  When the theory is in the conformal window, the IR $U(1)_R$ charge of the chiral multiplets is $r = (N_f - N_C)/N_F$.  When the gauge group is $SU(2),$ the fundamental and anti-fundamental representation are isomorphic, so it is equivalent to think of a theory with $2 N_f$ doublets of $SU(2).$  The number of $SU(2)$ doublets is necessarily even because of a global anomaly \cite{Witten:1982fp}.  When the gauge group is $SU(2),$ the global symmetry enhances to $SU(2 N_f) \times U(1)_R.$

The quarks $Q^i_a, i = 1, \dots 2 N_f$ are chiral superfields transforming in the fundamental representation of $SU(2 N_f)$ and $a$ is the color index for $SU(2).$  The gauge invariant mesons are chiral superfields $M^{ij}$ that are color singlets obtained from
$$M^{ij} = \epsilon^{ab} Q^i_a Q^j_b.$$  The mesons transform in the anti-symmetric $\wedge^2 (\mathbf{2n})$ representation of $SU(2N_f).$
The classical moduli space of vacua is parametrized by the space of possible expectation values of the mesons $M^{ij}$ subject to the constraint
$$M \wedge M = 0,$$
or equivalently
$$\epsilon_{i_1 j_1 i_2  j_2 \dots i_n j_n} M^{i_1 j_1} M^{i_2 j_2} = 0.$$
These equations imply that $M$ has rank at most two.  Viewing the mesons $M^{ij}$ as coordinates on projective space $\P(\wedge^2 (\bold{2n}))$, the above equations are the Pl\"ucker relations that describe the embedding of the Grassmannian $\Gr(2, 2n)$ as an algebraic subvariety of $\P(\wedge^2 (\bold{2n})).$  However, the space of expectation values of the meson fields is not projectivized.  Therefore, the classical moduli space of vacua is the affine cone over the Grassmannian $\Gr(2, 2n).$

%----------------------------------------------------------
\section{Local operators from polyvector fields}
%----------------------------------------------------------
We view the low energy effective theory of a four-dimensional theory as an $\mathcal{N}=1$ supersymmetric nonlinear sigma model from Minkowski space or $\R \times S^3$ to $\mathcal{M},$ the quantum moduli space of vacua.  Similar to the B-model in two dimensions, local BPS operators arise from polyvector fields on $\mathcal{M}.$  Recall that the B-model with target space $B$ has 
$$\bigoplus_{p,q} H^{q}(B, \wedge^p T^{1,0} B)$$
as its space of local observables \cite{Witten:1991zz}.
We will consider the case when the moduli space $\M$ can be described as an affine cone over a base $B$.
Then the local operators arise from the polyvector fields
$$\bigoplus_{p,q,k} H^{q}(B, \wedge^p T^{1,0} B \otimes \L^k)$$
where $\mathcal{L}$ is a line bundle over $\M.$  In the examples we consider of algebraic varieties embedded in projective space, $\mathcal{L}$ is the pull-back of $\O(1)$ on the ambient projective space.  This picture of local operators is described in more detail by Beasley and Witten \cite{Beasley:2004ys}.

In the case that $B$ is a homogeneous space, we can evaluate the relevant cohomology groups using a generalization of the Borel-Weil-Bott theorem.  The first step is to express the tangent bundle of $B$ as a homogeneous bundle.  Although not strictly necessary for the logical development, we make a brief detour to explain a physical description of the tangent bundle to the Grassmannian in $SU(2)$ SQCD.

%----------------------------------------------------------
\paragraph{Homogeneous bundles and $SU(2)$ SQCD}
%----------------------------------------------------------
The Grassmannian $\Gr(k, n)$ can be viewed as the space of complex $k$-planes $\Lambda \cong \bC^k$ in $\bC^{n}.$  The Grassmannian has a natural vector bundle called the {\it universal subbundle} $S \rightarrow G(k,n),$ which is the subbundle of
$\bC^n \times G(k,n)$ whose fiber at each point $\Lambda \subset G(k,n)$ is the subspace $\Lambda \subset \bC^n$.
The {\it universal quotient bundle} $Q \rightarrow G(k,n)$ is the quotient bundle
$Q = \bC^n/S$.  The subbundle and the quotient bundle fit into the {\it tautological exact sequence} 
$$0 \rightarrow S \rightarrow \bC^n \rightarrow Q \rightarrow 0$$
on the Grassmannian $\Gr \cong G(k,n).$
The tangent bundle of the moduli space $\M$ arising from fluctuations about a generic point on $\M$ recovers the algebraic geometry description of the tangent bundle as 
$$T_{\Gr} \cong \Hom_{\Gr}(S, Q).$$

Alternatively, $\M_{cl}$ can be described by the fluctuations about a fixed supersymmetric vacuum.  Up to gauge and global symmetry transformations, the classical moduli space of supersymmetric vacua has the form
$$Q^i_a =
\begin{pmatrix}
v \; & 0 \\
0 \; & v \\
0 \; & 0 \\
\vdots \; & \vdots \\
0 \; & 0
\end{pmatrix}
$$
where $v$ is an arbitrary complex number.  Unlike the gauge invariant description in terms of the mesons $M^{ij},$ this description explicitly depends on a choice of gauge.
When $v$ is non-zero, the expectation values of $Q^i_a,$ break the global symmetry group from
$SU(2n)$ to the subgroup $SU(2) \times SU(2n - 2).$
The gauge group is completely Higgsed.  We recover that the moduli space of vacua is the quotient space
$$\Gr(2,2n) \cong \frac{U(2n)}{U(2) \times U(2n-2)}.$$
The massless fluctuations of the quarks $Q^i_a$ about the vacuum transform in representations of the unbroken gauge group and are listed in table \ref{tab:fluc}.
\begin{table}[htp]
\begin{center}
\begin{tabular}{ccc}
& $SU(2)$ & $SU(2n-2)$ \\
$\Phi^s_c$ & $\mathbf{2}$ & $\bold{2n - 2}$ \\
$\Phi$ & $\bold{1}$ & $\bold{1}$
\end{tabular}
\end{center}
\caption{Massless fluctuations of quarks in $SU(2)$ SQCD}
\label{tab:fluc}
\end{table}
The field $\Dbar_{\dot{\alpha}} \Phibar$ represents a tangent vector to the moduli space $\M.$  Using the K\"ahler metric, it can be converted to a holomorphic one-form on $\M.$  It transforms in the bundle $T_{\Gr} \cong \Hom_{\Gr}(S, Q).$

%----------------------------------------------------------
\paragraph{Index of polyvector fields}
%----------------------------------------------------------
The description of the tangent bundle to the moduli space as the homogeneous bundle $T_{\Gr} \cong \Hom_{\Gr}(S, Q)$ is precisely what is needed in computation of the space of polyvector fields using a generalization of the Borel-Weil-Bott theorem.  This theorem describes the polyvector fields as representations of the global symmetry group.  We typically list only the dimensions of the representations, but stress that the actual representations can easily be determined.

Let $\mathcal{L}$ by the line bundle on $B,$ which is the restriction of $\O(1)$ of the ambient projective space.  Then for any coherent sheaf $\mathcal{F}$ on $X$ we denote $\mathcal{F}(m) = \mathcal{F} \otimes \mathcal{L}^{\otimes m}$.  The Euler character $\chi(\mathcal{F}(m))$ is a polynomial in $m$ called the {\it Hilbert polynomial}.  For $m \gg 0,$ these polynomials precisely count the contributions of local operators.  However, for small $m,$ some of the cohomology classes are not actually realized as local operators.  The unitarity bound $E \ge \frac{3}{2} r + 2 j _2$ excludes these cohomology elements from representing local operators.  However, we will see later that they can correspond to Beasley-Witten F-terms. 
%----------------------------------------------------------
\paragraph{Polyvector fields on Severi varieties}
%----------------------------------------------------------
We define the alternating sum of Euler characters of polyvector fields on the Severi varieties to be
$$\chi(t) = \sum_{m = 2j}^{\infty} \sum_{j=0}^{\dim B} (-1)^j \chi(\wedge^j TB(-3) \otimes \O(m)) t^{2m/3}.$$
The restriction of $m \ge 2j$ implements the unitarity bound.

For Severi varieties, this alternating sum of Euler characteristics is
\begin{equation}
\chi(t) = \frac{(1-t^{4/3})^{\dim(V_{\lambda})}}{(1-t^{2/3})^{\dim(V_{\lambda})}(1-t)} - \frac{t^{2} - t^{2 \dim(V_{\lambda})/3} }{(1-t^2)} 
\label{eq:chi}
\end{equation}
where $\dim(V_{\lambda})$ is the number of chiral multiplets.
In terms of the plethystic logarithm
$$\Exp^{-1}[\chi(t)(1 - t^2) + t^2 - t^{2 \dim(V_{\lambda})/3}] = \dim(V_{\lambda})t^{2/3} - \dim(V_{\lambda}) t^{4/3}.$$

%----------------------------------------------------------
\section{Superconformal index from the moduli space}
%----------------------------------------------------------
In this section, we will explain how the local operators constructed from the moduli space of vacua can be related to the superconformal index.
First, we review the definition of the superconformal index following \cite{Dolan:2008qi}.  Then we will propose a formula for the superconformal index in terms of the index of polyvector fields on the moduli space of vacua.

%----------------------------------------------------------
\paragraph{Superconformal index review}
%----------------------------------------------------------
The superconformal index $\mathcal{I}(t,y,h)$ is defined as
$$\mathcal{I}(t,y, h_a) = \Tr_{\mathcal{H}} \;  (-1)^F t^{2(E + j_2)} x^{2 j_1}  h_a^{F_a} $$
where the trace is taken over the Hilbert space $\mathcal{H}$ of the theory quantized on $\R \times S^3.$
By the operator-state correspondence, the index can also be viewed as a trace over the Hilbert space of local operators.
In the formula $F$ is the fermion number, $j_{1,2}$ are the left and right spins, and $E$ is the operator scaling dimension.
If the theory has flavor symmetries, then we can introduce an equivariant index with flavor fugacities $h_a$ for the flavor symmetry charges $F_a.$
For a theory with a UV Lagrangian description with gauge group $G,$ and chiral multiplets $\Phi_i$ in the representation $R_{G,i}$ of $G$ the index is given by the following integral
$$\mathcal{I}(t,y, h) = \int_{G} d \mu(g) \Exp \left( \sum_{n=1}^{\infty} i(t^n, y^n, h^n,g^n) \right)$$
over $G$ with respect to the Haar measure  $d \mu(g)$.  The integrand is the plethystic exponential of the single letter index
\begin{align}
i(t,y,h,g) & = \frac{2t^2 - t \chi_2(y)}{(1 - ty)(1-t y^{-1})} \chi_{adj}(g)  \\
& + \sum_i \frac{t^{r_i} \chi_{R_{F,i}}(h) \chi_{R_{G,i}}(g) - t^{2-r_i} \chi_{\overline{R}_{F,i}}(h) \chi_{\overline{R}_{G,i}}(g)}{(1 - ty)(1-t y^{-1})}
\end{align}
where
$ \chi_2(y) = y + y^{-1}$ and $r_i$ are the $U(1)_R$ charges of the chiral multiplets.  The first contribution is from the vector multiplet accounts and $\chi_{adj}$ is the adjoint character of the gauge group.  Similarly, $\chi_{R_{G,i}}(g)$ are the characters of the representations $R_{G,i}$ of the gauge group
 and $\chi_{R_{F,i}}(h)$ are the characters of the representations of the chiral multiplets in the flavor symmetry group.
The plethystic exponential is defined by
$$f(t) = \sum_{n \ge 1} a_n t^n \mapsto \Exp[ f(t)] = \prod_{n \ge 1} \frac{1}{(1 - t^n)^{a_n}}$$
and can similarly be extended to several variables.

%----------------------------------------------------------
\paragraph{Superconformal index from the moduli space}
%---------------------------------------------------------- 
All of the short multiplets contributing to the $\mathcal{N} = 1$ superconformal limits can be viewed as special cases of the $\mathcal{C}$ multiplets with possibly unphysical spins and $r$-charges.
A $\mathcal{C}_{r(j_1,j_2)}$ multiplet satisfies the shortening condition $\Delta = 2 + 2 j_1 + \frac{3}{2} r$ \cite{Rastelli:2016tbz}.
Conserved currents reside in $\widehat{\mathcal{C}} _{0(1/2,1/2)}$ multiplets and have $\Delta = 3.$
Defining $\widetilde{r} = 2 j_1 + r$, the contribution of a $\mathcal{C}_{r(j_1,j_2)}$ multiplet to the index is
$$\mathcal{I}_{[\widetilde{r},j_2]}(t,y) = (-1)^{2j_2 +1} \frac{t^{\widetilde{r} + 2} \chi_{j_2}(y)}{(1-t y)(1-t y^{-1})}.$$
For a fixed set of quantum numbers $(\widetilde{r},j_2)$, the {\it net degeneracy} $ND[\widetilde{r},j_2]$ is defined to be
the number of short multiplets with integer spin minus the number of short multiplets with half integer spin.
The net degeneracy can be determined from the superconformal index using the relation \cite{Beem:2012yn}
$$(1-t y)(1-t y^{-1}) \mathcal{I}(t,y)  = \sum_{\widetilde{r}, j_2} ND[\widetilde{r}, j_2] t^{\widetilde{r} + 2} \chi_{j_2}(y).$$
For quantum numbers $[\widetilde{r},j_2]$, the net degeneracy is
\begin{align*}
ND[0,0] & = \# \text{marginal operators} - \# \text{conserved currents} \\
             & = H^0(B, \mathcal{L}) - H^0(B, TB),
\end{align*}
which gives the dimension of the conformal manifold \cite{Green:2010da}.
In general we expect contributions not just from the tangent bundle $TB$ but from all exterior powers $\wedge^{\bullet} TB.$
However, these operators are no longer sufficient to account for all of the contributions to the net degeneracy.  From experience with holographic duality, it is natural to expect that the operators arising from polyvector fields generate a Fock space of operators that contribute to the index.

%----------------------------------------------------------
\paragraph{Superconformal index from polyvectorfields}
%---------------------------------------------------------- 

Given the index $\chi(t)$ of polyvector fields we propose that the superconformal index $\mathcal{I}(t,y)$ can be expressed as
$$(1-t y)(1-t y^{-1}) \Exp^{-1}[ \mathcal{I}(t,y)]  = \Exp^{-1}[\chi(t)(1 - t^2) + t^2] + \dots,$$
where the correction terms arise from massless matter at the singularity.

For the cones over the Severi varieties using $\chi(t)$ from Equation \eqref{eq:chi}
$$(1-t y)(1-t y^{-1}) \Exp^{-1}[ \mathcal{I}(t,y)] =  \dim(V_{\lambda})t - \dim(V_{\lambda}) t^2$$
We therefore recover the superconformal index
$$\mathcal{I}(t,y) = \Exp \left[\frac{\dim(V_{\lambda})t^{2/3} - \dim(V_{\lambda}) t^{4/3} }{(1-t y)(1-t y^{-1})} \right]. $$
In general, there is not an exact match.
For SQCD with four flavors,
\begin{align*}
\Exp^{-1} & [ \mathcal{I}(t,y)](1-t y)(1-t  y^{-1}) \\
  & =  \frac{1}{2} \mathbf{56} t - \mathbf{133} t^2+\left(\mathbf{912} -\mathbf{1}\chi_2(y) \right) t^3+\left(-(\mathbf{8645 + 133})+
  \mathbf{56}\chi_2(y) \right) t^4 + \O(t^{5})
\end{align*}
while the index constructed from polyvector fields is
\begin{align*}
\Exp^{-1}[\chi(t)(1 - t^2) + t^2] & = \frac{1}{2} \mathbf{56} t - \mathbf{133} t^2 + \mathbf{912} t^3 - (\mathbf{8645 + 133}) t^4 + \O(t^{5}).
\end{align*}
The two indices are remarkably close, but already differ by differ at their $t^3$ term.  The extra massless degrees of freedom at the origin of the moduli space are responsible for the failure of equality of the two indices.  By taking the difference of the two indices, we can isolate the new extra massless degrees of freedom at the origin of moduli space.  Alternatively, it is an intriguing problem to take into account polyvector fields localized at the origin to give an exact formula for the superconformal index.  In the next section we will explain the computation of these indices in more detail. 

%----------------------------------------------------------
\section{Polvectorfields on moduli spaces of vacua}
%----------------------------------------------------------
%----------------------------------------------------------
\paragraph{Polyvector fields on $\P^2  \subset \P^5$ and $\P^2 \times \P^2 \subset \P^8$}
%----------------------------------------------------------
The polyvector fields on the Veronese embedding of $\P^2$ in $\P^5$ are listed in table \ref{pvfieldsver}.  They polyvector fields that are sections of $\O(m)$
are in the representation $\Sym^{2m} \mathbf{3}$ of $SU(3).$  The sections of $TX(m)$ are in $[2m - 1, 1]$ representation of $SU(3)$ in the notation of LiE.  The sections of $\wedge^2 TX(m)$ transform in the $\Sym^{2m + 3} \mathbf{3}$ representations of $SU(3).$
\begin{table}[h]
\begin{center}
\begin{tabular}{c||c|c|c|c||c|}
dim & $\O_X$ & $TX$ & $\wedge^2 TX$   \\
\hline
0 & 1 & 0 & 0 & 1 \\
1 & 6 & 0 & 0 & 6 \\
2 & 15 & 0 & 0 & 15 \\
3 & 28& -8 & 0 & 20 \\
4 & 45 & -24 & 0 & 21 \\
5 & 66 & -48 & 3 & 21 \\
6 & 91 & -80 & 10 & 21 \\
\hline
\end{tabular}
\end{center}
\caption{$\chi$ of polyvector fields on $\P^2  \subset \P^5$.}
\label{pvfieldsver}
\end{table}%
For the Segre embedding of $\P^2 \times \P^2 \subset \P^8$ the polyvector fields transform in representations of $SU(3) \times SU(3)$
and are listed in table \ref{pvfieldscp22}.
\begin{table}[h]
\begin{center}
\begin{tabular}{c||c|c|c|c|c|c||c|}
dim & $\O_X$ & $TX$ & $\wedge^2 TX$ & $\wedge^3 TX$  & $\wedge^4 TX$  \\
\hline
0 & 1 & 0 & 0 & 0 & 0 & 1\\
1 & 9 & 0 & 0 & 0 & 0 & 9 \\
2 & 36 & 0 & 0 & 0 & 0 & 36\\
3 & 100 & -16 & 0 & 0 & 0 & 84 \\
4 & 225 & -90 & 0 & 0 & 0 & 135 \\
5 & 441 & -288 & 9 & 0 & 0 & 162 \\
6 & 784 & -700 & 84 & 0 & 0 & 168 \\
7 & 1296 & -1440 & 315 & 0 & 0 & 171 \\
8 & 2025 & -2646 & 828 & -36 & 0 & 171 \\
9 & 3025 & -4480 & 1785 & -160 & 1 & 171 \\
10 & 4356 & -7128 & 3384 & -450 & 9 & 171 \\
\hline
\end{tabular}
\end{center}
\caption{$\chi$ of polyvector fields on $\P^2 \times \P^2 \subset \P^8$.}
\label{pvfieldscp22}
\end{table}%
%----------------------------------------------------------
\paragraph{Polyvector fields on $Gr(2,6)$}
%----------------------------------------------------------
We can apply Borel-Weil-Bott theorem as reviewed in Appendix \ref{sec:BWB}.  Several of the cohomology groups are listed in table \label{pvfieldsgr26} in Appendix \ref{sec:PVF}.
%----------------------------------------------------------
\paragraph{Polyvector fields on the Cayley plane}
%----------------------------------------------------------
\begin{table}[h]
\begin{center}
\begin{tabular}{c||c|c|c|c|c|c||c|}
dim & $\O_X$ & $TX$ & $\wedge^2 TX$ & $\wedge^3 TX$  & $\wedge^4 TX$ & $\wedge^5 TX$  & \\
\hline
0 & 1 & 0 & 0 & 0 & 0 &  0 & 1\\
\hline
1 & 27 & 0 & 0 & 0 & 0 &  0 & 27 \\
\hline
2 & 351 & 0 & 0 & 0 & 0 &  0  & 351 \\
\hline
3 & 3003 &  -78 & 0 & 0 &  0 & 0 & 2925 \\
\hline
4 & 19305 &  -1728 & 0 & 0 & 0 & 0  & 17577\\
\hline
5 & 100386 & -19305 & 0 &  0 & 0 & 0  & 81081\\
\hline
6 & 442442 & -146432 & 2925 &  0 & 0 & 0 & 298935\\
\hline
7 & 1706562 & -853281 & 51975 & 351 &  0  & 0 & 905607 \\
\hline
8 & 5895396 & -4088448 & 494208 & 0 &  0 & 0  & 2301156\\
\hline
9 & 18559580 & -16812796 & 3309696 & -70070 & -650 & 0 & 4985760\\
%\hline
%10 & 53965548 & -61102080 & 17453475 & -967680 & -7371 & 0 & 0 & 0 & 9341892\\
%\hline
%11 & 146477916 & -200443464 & 77026950 & -7757100 & 34398 & 351 & 0 & 0 & 15339051\\
%\hline
%12 & 374332452 & -603043584 & 295609600 & -45741696 & 634998 & 0 & 0 & 0 & \\
\hline
\end{tabular}
\end{center}
\caption{$\chi$ of polyvector fields on $OP_2$.}
\label{pvfieldsOP2}
\end{table}%

\begin{figure}[h]
\begin{center}
\begin{tikzpicture}
\path (0,0) node[draw,shape=circle] (v1) {} node[below,yshift = -1ex]{$\alpha_1$};
\path (1,0) node[draw,shape=circle] (v3) {} node[below,yshift = -1ex]{$\alpha_3$};
\path (2,0) node[draw,shape=circle] (v4) {} node[below,yshift = -1ex]{$\alpha_4$};
\path (3,0) node[draw,shape=circle] (v5) {} node[below,yshift = -1ex]{$\alpha_5$};
\path (4,0) node[draw,shape=circle] (v6) {} node[below,yshift = -1ex]{$\alpha_6$};
\path (2,1) node[draw,shape=circle] (v2) {} node[above,yshift = 1ex]{$\alpha_2$};
\draw (v1)--(v3)--(v4)--(v5)--(v6);
\draw (v4)--(v2);
\end{tikzpicture}
\caption{$E6$ Dynkin Diagram} 
\label{fig:e6}
\end{center}
\end{figure}
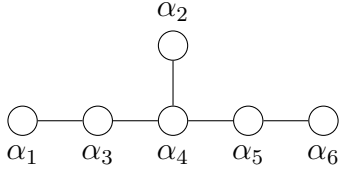
The Cayley plane $X = \mathbb{OP}^2 \cong E_6/P(\alpha_1)$
where $P(\alpha_1)$ is the parabolic subgroup corresponding to the root $\alpha_1$ of $E_6.$ The Dynkin diagram and labelling of roots for $E_6$ is shown in Figure \ref{fig:e6}.
The root $\alpha_1$ is dual to the weight $\omega_1.$
%Similar to $Gr(2,6),$ we find that the alternating sum of Euler characters of polyvector fields on $\mathbb{OP}^2$ is given by 
%$$\frac{(1-t^4)^{27}}{(1-t^6)(1-t^2)^{27}} - \frac{t^6}{(1-t^6)} $$
The first three tensors powers of tangent bundle of $X = \mathbb{OP}^2$ are \cite{MR3293722}
$$\cT_X \cong \E_{\omega_2}, \qquad \wedge^2 \cT_X \cong \E_{\omega_4}, \qquad \wedge^3 \cT_X \cong \E_{\omega_3 + \omega_5}.$$
The Euler characteristics of polyvector fields on the Cayley plane are listed in table \ref{pvfieldsOP2}.
%----------------------------------------------------------
\paragraph{Hilbert series and free resolutions}
%----------------------------------------------------------
A striking feature of the table of local operators, is that there is a diagonal of zeros.  This motivates the idea of slicing the table diagonally instead of horizontally or vertically.  The first non-trivial diagonal is closely related to a free resolution of the structure sheaf.  For the Veronese embedding of $\P^2$ in $\P^5$ we have the (graded) free resolution
$$\O \rightarrow \O_{\P^5}(-4)^{\oplus 3} \rightarrow \O_{\P^5}(-3)^{\oplus 8} \rightarrow \O_{\P^5}(-2)^{\oplus 6} \rightarrow \O_{\P^5} \rightarrow \O_{\P^2} \rightarrow 0$$
of $\O_{\P^2}$ \cite{MR1818055}.  From the free resolution of $\O_{P^2}$, we can determine the Hilbert series of the Veronese embedded $\P^2,$
$$H(\P^2; t) = \frac{1 - 6 t^2 + 8 t^3 - 3 t^4}{(1-t)^6}$$
For Segre embedded $\P^2 \times \P^2 \hookrightarrow \P^8$ the free resolution is constructed in \cite{MR3154839} using the methods of \cite{MR2384793}.  The free resolution of the Grassmannian $Gr(2,6)$ is described in \cite{MR554859, MR1285374}.  For the complex Cayley plane $\mathbb{OP}^2$ the free resolution is given in Lemma 7.2 of \cite{MR3439089}.
\begin{table}[htp]
\begin{center}
{\tabulinesep=0.5mm
\begin{tabu}{|c|c|}
\hline
 $X$ & Hilbert Series $H(X; t)$\\
\hline
$\P^2 \subset \P^5$ & $\frac{1 + 3 t}{(1 - t)^3}$ \\
\hline
$\P^2 \times \P^2 \subset \P^8$ & $\frac{1+4 t+t^2}{(1-t)^5}$ \\
\hline
$\Gr(2,6)$ & $\frac{1 + 6 t + 6 t^2 + t^3}{(1 - t)^9}$ \\
\hline
$\mathbb{OP}^2$ & $\frac{1+10 t+28 t^2+28 t^3+10 t^4+t^5}{(1-t)^{17}}$ \\
\hline
\end{tabu}
}
\end{center}
\label{tab:HS}
\caption{Hilbert Series of Severi varieties}
\end{table}%

The Hilbert series of the Severi varieties is also
$$H(X; t) = \sum_{k \ge 0} \dim V_{k \lambda} t^k$$
where $\lambda$ is the weight of the defining representation $V_{\lambda}$ of the corresponding Wess-Zumino model.
Then $V_{k \lambda}$ is the representation with weight $k \lambda.$  This representation geometrically is the space of sections of the line bundle $\O(k),$ 
$$H^0(X, \O(k)) = V_{k \lambda}.$$
The Hilbert series of the Severi varieties are listed in Table \ref{tab:HS} using results from \cite{MR2906911}.
The Hilbert series can also be expressed in the form
$$H(X;t) = \frac{N(X;t)}{(1 - t)^{\dim(V_{\lambda})}}$$
For the first three Severi varieties, the terms in the numerator $N(X;t)$ determines its minimal free resolution.  It is natural to conjecture that this also holds for
$\mathbb{OP}^2.$  The numerators $N(X;t)$ of $Gr(2,6)$ and $\mathbb{OP}^2$ are listed in Table \ref{tab:HSnum}.  The corresponding betti tables of the minimal free resolutions are 
\[
\begin{bmatrix}
  1  &  -  &  -  &  -  &  -  &  -  &  -    \\
  -  &  15  &  35  &  21  &  -  &  -  &  -    \\
  -  &  -  &  -  &   21  &  35  &  15  &  -     \\
  -  &  -  &  -  &  -  &  -  &  -  &   1    \\
\end{bmatrix}
\]
for $Gr(2,6)$ and
\setcounter{MaxMatrixCols}{20}
\[
\begin{bmatrix}
  1  &  -  &  -  &  -  &  -  &  -  &  -  &  -  &  -  &  -  &  -  \\
  -  &  27  &  78  &  -  &  -  &  -  &  -  &  -  &  -  &  -  &  -  \\
  -  &  -  &  -  &  351  &  650  &  351  &  -  &  -  &  -  &  -  &  -  \\
  -  &  -  &  -  &  -  &  -  &  351  &  650  &  351  &  -  &  -  & - \\
  -  &  -  &  -  &  -  &  -  &  -  &  -  &  -  &  78  &  27  &  -  \\    
  -  &  -  &  -  &  -  &  -  &  -  &  -  &  -  &  -  &  -  &  1  \\ 
\end{bmatrix}
\]
for $\mathbb{OP}^2$.
%\cite{MR3439089}
\begin{table}[htp]
\begin{center}
\begin{tabular}{|c|c|}
\hline
$X$ & $N(X; t)$ \\
\hline
\hline
%$\P^2 \subset \P^5$ & $1 - 6 t^2 + 8 t^3 - 3 t^4$ \\
%\hline
%$\P^2 \times \P^2 \subset \P^8$ & $1 - 9 t^2 + 16 t^3 - 9 t^4 + t^6$ \\
%\hline
$\Gr(2,6)$ &  $1-15 t^2+35 t^3-21 t^4-21 t^5+35 t^6-15 t^7+t^9$ \\
\hline
$\mathbb{OP}^2$ &  $1-27 t^2+78 t^3-351 t^5+650 t^6-351 t^7-351 t^8+650 t^9-351 t^{10}+78 t^{12}-27 t^{13}+t^{15}$ \\
\hline
\end{tabular}
\end{center}
\label{tab:HSnum}
\caption{Hilbert series numerators for $Gr(2,6)$ and $\mathbb{OP}^2$}
\end{table}%
Intriguingly the dimension of the cohomology groups along one of the diagonals precisely matches the number of  simplices in the conjectured minimal triangulation of the projective plane $\mathbb{AP}^2$ \cite{MR3038783}\footnote{The author would like to thank Sergey Galkin for this observation.}.
%----------------------------------------------------------
\section{The $E_7$ surprise revisited}
%----------------------------------------------------------
Dimofte and Gaiotto argued that four dimensional SQCD with eight chiral multiplets defines an $E_7$ superconformal invariant boundary condition for a five dimensional theory of 28 free hypermultiplets \cite{Dimofte:2012pd}.  One strong hint for the $E_7$ symmetry is that there are 72 dual descriptions of the SQCD theory corresponding to elements of the quotient of Weyl groups $W(E_7)/W(A_7)$ \cite{Spiridonov:2008zr}.  We will find that the index of polyvector fields on the moduli space of vacua of  SQCD with eight chiral multiplets  can be expressed in terms of the Hilbert series of the Freudenthal variety $E_7/P(\alpha_7)$.  As a consequence, our conjectured form of the index in terms of polyvector fields also predicts the $E_7$ enhancement of the $4D/5D$ superconformal index.

Similarly three dimensional SQCD with six chiral multiplets defines an $SO(12)$ superconformal invariant boundary condition for 16 four dimensional free hypermultiplets \cite{Dimofte:2012pd}.  Again we will see that the index of polyvector fields can be expressed in terms of the Hilbert series of the spinor variety $\mathbb{S}_{12}.$

Quite remarkably the enhanced symmetry groups $SO(12)$ and $E_7$ occur in the third row of the Freudenthal magic square.  Their corresponding Landsberg-Manivel projective geometries are the spinor variety $\mathbb{S}_{12}$ and the Freudenthal variety.
%----------------------------------------------------------
\paragraph{Polyvector fields on $Gr(2,8)$}
%----------------------------------------------------------
\begin{table}[h]
\begin{center}
\begin{tabular}{c||c|c|c|c|c|c||c|}
dim & $\O_X$ & $TX$ & $\wedge^2 TX$ & $\wedge^3 TX$  & $\wedge^4 TX$ & $\wedge^5 TX$  & \\
\hline
0 & 1 & 0 & 0 & 0 & 0 &  0 & 1\\
\hline
1 & 28 & 0 & 0 & 0 & 0 & 0 & 28 \\
\hline
2 & 336 & -63 & 0 & 0  & 0 &  0 & 273 \\
\hline
3 & 2520 & -1280 & 36 & 0 & 0 & 0 & 1276 \\
\hline
4 & 13860 & -12474 & 1890 & 0 &  0 & 0 & 3276 \\
\hline
5 & 60984 &  -80640 & 26180 & -1280 & 36 &  0 & 5280 \\
\hline
6 & 226512 & -396396  & 205920 & -30030 & 336 & -63 & 6279 \\
\hline
%7 & 736164 & -1596672 & 1149876 & -302848 & 19980 & 0 & 28 & 0 & 0 & 6528 \\
%\hline
\end{tabular}
\end{center}
\caption{$\chi$ of polyvector fields on $Gr(2,8)$.}
\label{pvfieldsGr28}
\end{table}%

The moduli space of vacua of four dimensional $SU(2)$ SQCD with four flavors is an affine cone over the Grassmannian $Gr(2,8).$  The theory does not have a dual magnetic description as a Wess-Zumino model.  Instead it has a rich duality web.  Nevertheless, we will see that we can still determine a rich set of operators from its moduli space of vacua.  The calculation of cohomologies of local operators is practically identical to the case of $Gr(2,6)$ again using the Borel-Weil-Bott theorem.
The first few cohomology groups are shown in table \ref{pvfieldsGr28} and more cohomology groups are listed in table \ref{pvfieldsGr28C} in the appendix.  The first few terms in the betti table of the minimal free resolution are
\[
\begin{bmatrix}
  1  &  -  &  -  &  -  &  -  &  -  &  \dots    \\
  -  &  70  &  420  &  945  &  924  & 330  &  \dots    \\
  -  &  -  &  -  &   1176  &  7350  &  19980  &  \dots     \\
\end{bmatrix}
\]
and as representations of $SU(8)$ the betti table is
\[
\begin{bmatrix}
  V_0  &  -  &  -  &  -  &  -  &    \dots    \\
  -  &  V_{\omega_4}  &  V_{\omega _ + \omega_5}  &  V_{2 \omega_1 + \omega_6} &  V_{3 \omega_1 + \omega_7}    &  \dots    \\
  -  &  -  &  -  &   V_{2 \omega_5}  &  V_{\omega_1 + \omega_5 + \omega_6}  &   \dots     \\
\end{bmatrix}
\]
where we have listed only the first few terms.

The r-charges of the chiral multiplets are $1/2$ so the alternating sum of polyvector fields takes the form
$$\chi(t) = \sum_{m = 0}^{\infty} \sum_{j=0}^{\dim B} (-1)^j \hat{\chi}(\wedge^j TB(-2) \otimes \O(m)) t^{2m/4}$$
The alternating sum of Euler characters of polyvector fields on $Gr(2,8)$ can be written as
$$\chi(t)  = \frac{-t^2 +t^{10} + 2 t^{14}}{1-t^2} + \frac{P(t)}{(1-t^2)}$$
where
$$P(t) = \left( 1 + 28 t + 273 t^2 + 1248 t^3 + 3003 t^4 + 4004 t^{5} + 3003 t^{6} + 
 1248 t^{7} + 273 t^{8} + 28 t^{9} + t^{10} \right).$$
The polynomial $P(t)$ is the numerator of the Hilbert series 
$$\frac{P(t)}{(1 - t^{1/2})^{28}}$$
of the Freudenthal variety $E_7/P(\alpha_7)$  \cite{MR2906911}
where $P(\alpha_7)$ is the parabolic subgroup corresponding to the root $\alpha_7$ of $E_7.$ 

Taking the plethystic logarithm of the Hilbert series, we find 
\begin{align*}
\Exp^{-1}[\chi(t)(1 - t^2) + t^2] & = 28 t - 133 t^2 + 912 t^3 - 8778 t^4 + 93632 t^{5}  + \O(t^{6}) \\
& = \frac{1}{2} \mathbf{56} t - \mathbf{133} t^2 + \mathbf{912} t^3 - (\mathbf{8645 + 133}) t^4 + \O(t^{5})
\end{align*}
The exact result for the index is
\begin{align*}
\Exp^{-1} & [ \mathcal{I}(t,y)](1-t y)(1-t  y^{-1}) \\
  & =  \frac{1}{2} \mathbf{56} t - \mathbf{133} t^2+\left(\mathbf{912} -\mathbf{1}\chi_2(y) \right) t^3+\left(-(\mathbf{8645 + 133})+
  \mathbf{56}\chi_2(y) \right) t^4 + \O(t^{5}),
\end{align*}
which is remarkably close.
%----------------------------------------------------------
\paragraph{3D Index}
%----------------------------------------------------------
\begin{table}[h]
\begin{center}
\begin{tabular}{c||c|c|c|c|c|c|c|c|c||c|}
dim & $\O_X$ & $TX$ & $\wedge^2 TX$ & $\wedge^3 TX$  & $\wedge^4 TX$ & $\wedge^5 TX$ & $\wedge^6 TX$ & $\wedge^7 TX$ & $\wedge^8 TX$ & \\
\hline
0 & 1 & 0 & 0 & 0 & 0 & 0 & 0 & 0 & 0 & 1\\
\hline
1 & 15 & 0 & 0 & 0 & 0 & 0 & 0 & 0 & 0 & 15 \\
\hline
2 & 105 & -35 & 0 & 0 & 0 & 0 & 0 & 0 & 0 & 70 \\
\hline
3 & 490 & -384 & 21 & 0 & 0 & 0 & 0 & 0 & 0 & 127 \\
\hline
4 & 1764 & -2205 & 560 & 21 & 0 & 0 & 0 & 0 & 0 & 140 \\
\hline
5 & 5292 & -8960 & 4230 & -384 & -35 & 0 & 0 & 0 & 0 & 143 \\
\hline
6 & 13860 & -29106 & 19782 & -4515 & 105 & 15 & 0 & 0 & 0 & 141 \\
\hline
7 & 32670 & -80640 & 70070 & -24960 & 3003 & 0 & 0 & 0 & 0 & 143 \\
\end{tabular}
\end{center}
\caption{$\chi$ of polyvector fields on $Gr(2,6)$ with alternate grading for 3d SQCD}
\label{pvfieldsgr26b}
\end{table}%
Similarly the moduli space of vacua for three dimensional SQCD with three fundamental flavors is a cone over the Grassmannian $\Gr(2,6).$  However, unlike in four dimensional SQCD with three fundamental flavors, the R-charge assignments are different.  This means that the index of polyvector fields is different then the one computed in four dimensions also for $\Gr(2,6)$.
The index computed from \cite{Dimofte:2012pd} is
\begin{align*}
\mathcal{I}(q)_{3D/4D} = 1 & + 32 q^{1/2} + 462 q + 4256 q^{3/2} + 29271 q^2 + 164064 q^{5/2} \\
& + 789558 q^3 + 3372864 q^{7/2} + 13085623 q^4 + 46874080 q^{9/2} + O(q^5).
\end{align*}
After taking the plethystic logarithm, 
\begin{align*}
\Exp^{-1} & [ \mathcal{I}(q)](1-q^2) \\
& =  32 q^{1/2} - 66 q + 352 q^{3/2} - 2608 q^2 + 21152 q^{5/2} - 178124 q^3 \\
& + 1543904 q^{7/2} - 13683946 q^4 + 123259552 q^{9/2} +  O(q^{5})
\end{align*}

The alternating sum of Euler characteristics
 $$\chi(q) = \frac{P(q)}{1-q^2} - \frac{q}{1-q},$$
 where
$$P(q) = 1 + 16 q^{1/2} + 70 q + 112 q^{3/2} + 70 q^2 + 16 q^{5/2} + q^3$$
is the numerator of the Hilbert series of the spinor variety $\mathbb{S}_{12} \cong \text{OGr}_{+}(6,12).$
Our conjectured formula for the index in three dimensions is
$$(1- q^2) \Exp^{-1}[ \mathcal{I}(q)]  = \Exp^{-1}[P(q)] + \dots,$$
$$Exp^{-1}[P(q)] = 16 q^{1/2} - 66 q + 352 q^{3/2} - 2607 q^2 + 21120 q^{5/2} + O(q^3)$$
%----------------------------------------------------------
\paragraph{Chern classes}
%----------------------------------------------------------
All of the alternating sums of Euler characteristics $\chi(t)$ have power series expansions that asymptote to either a limiting value or a limiting cycle of values.  The limiting value (or the average value of the limiting cycle) can be expressed in terms of the Hirzebruch-Riemann-Roch theorem.  The result is that the limiting value is the integral of the top Chern class of a twist of the tangent bundle. 
From the moduli spaces of vacua in 4d $SU(2)$ SQCD with $N_c = 3,4$ we find
$$\int_{\Gr(2,6)} c_{\text{top}}(TX(-3)) = 10923$$
$$\int_{\Gr(2,8)} c_{\text{top}}(TX(-2)) = 6556 = \frac{1}{2} deg(E_7/P(\alpha_7)) + 1.$$
The Chern classes of the Severi varieties are closely related to the Euler characteristics of their projective duals.
Similarly in three dimensions we find
$$\int_{\Gr(2,6)} c_{\text{top}}(TX(-2)) = 143 = \frac{1}{2} deg(\mathbb{S}_{12}).$$
These relations provide a simple mathematical realization of the $E_7$ and $SO(12)$-surprises.
%----------------------------------------------------------
\section{Projective duality and Beasley-Witten F-terms}
%----------------------------------------------------------
Beasley and Witten \cite{Beasley:2004ys} considered multi-fermion F-terms arising from cohomology classes on the moduli space of vacua.
By a direct computation in the generalized Wess-Zumino model with a cubic superpotential, Beasley and Witten argue that there are
multi-fermion F-terms in $SU(N_c)$ SQCD with $N_f$ flavors arising from sections of $$\Omega_X^{n-1} \otimes \wedge^{n-1} TX \otimes \O(-n)$$
where $n = N_f - N_c + 1.$
For both $\Gr(2,6)$ and $\mathbb{OP}^2$ there is a cohomology class \cite{MR3299729}
$$H^2(X, \wedge^2 TX(-3)) \cong \bC.$$ 
This appears to be a great coincidence from a direct calculation in the sigma model using the Borel-Weil-Bott theorem.  However there is a natural explanation in terms of projective duality.
For a Fano base $B,$ the Fano index $r$ is defined by $K_{B} = -r H$ where $K_B$ is the canonical class of $B$ and $H$ is the generator of the Picard group of $B.$  The co-index $c$ is the difference between the dimension and index of $B$, namely $c = \dim(B) - r.$  When the degree of the projective dual hypersurface is equal to the co-index minus one, then Lemma 4.2 in \cite{MR3299729} shows that the existence of a non-trivial cohomology class 
$$H^{c-2}(\Sigma, \wedge^{c-2} T\Sigma(-c+1)) \cong \bC$$  
using projective duality.  It is natural to surmise that the computations using projective duality are secretly the same as the computations at a generic point on the moduli space of vacua of a Landau-Ginzburg model.
%----------------------------------------------------------
\section{Schouten-Nijenhuis bracket and the OPE}
%----------------------------------------------------------
Given two vector fields $X$ and $Y$ on $B$, their Lie bracket $[X,Y]$ is another vector field on $B$.  There is an extension of the Lie bracket to polyvector fields known as the Schouten-Nijenhuis bracket.  We write the multivector fields on $B$ as $\mathfrak{X}^{\bullet}(B) = \Gamma(\wedge^{\bullet}TB)$.  A multivector field $P$ in $\Gamma(\wedge^{p}TB)$ has degree $|P| = p$.  Given multivector fields $P, Q \in \mathfrak{X}^{\bullet}(B)$ of degrees $|P| = p$ and $|Q| = q$, their Schouten-Nijenhuis bracket is of degree $|p| + |q| - 1.$  Under the Schouten-Nijenhuis bracket, the multivector fields on $B$ form a grade supercommutative ring.  Since local operators are identified with multivector fields it is natural to ask for a physical interpretation of their bracket.

The Schouten-Nijenhuis bracket on multivector fields has a direct translation into the operator product expansion (OPE) of local operators \cite{Costello15}.  The simplest example is the ordinary Lie bracket of two vector fields.  Vector fields (with the appropriate R-charge) correspond to conserved currents.  In 4d $\mathcal{N} = 1$ supersymmetry  conserved currents $j_{\mu}$ have scaling dimension $\Delta = 3$ and reside in supersymmetry multiplets $\mathcal{J}$ with leading component $\Delta = 2$.  The conserved current superfield has the following expansion
$$\mathcal{J}(z) = J(x) + i \theta j(x) - i \overline{\theta} \overline{j}(x) - \theta \sigma^{\mu} \overline{\theta}  j_{\mu}(x)$$
The OPE of two conserved current superfields is  \cite{Fortin:2011nq}
$$J_a(x) J_b(0) = \tau \frac{\delta_{ab} 1}{16 \pi^4 x^4} + \frac{k d_{abc}}{\tau} + \frac{J_c(0)}{16 \pi^2 x^2} - f_{abc} \frac{x^{\mu} j^c_{\mu}(0)}{24 \pi^2 x^2} + \dots$$
where $k$ is the corresponding t'Hooft anomaly and $f_{abc}$ are the structure constants of the symmetry algebra of the currents.  It is natural to identify these structure constants with the bracket of the corresponding vector fields.

It is also instructive to consider the description of the Schouten-Nijenhuis bracket in terms of the holomorphic twist of a 4d supersymmetric sigma model.
Given two local operators $\mathcal{O}_a$ and $\mathcal{O}_b$, we can place $\mathcal{O}_b$ at the origin and consider $\mathcal{O}_a^{(1)}$ arising from holomorphic descent of holomorphic descent of $\mathcal{O}_a$.  We can then integrate $\mathcal{O}_a^{(1)}$ over the cycle $H^1(\bC^2 \setminus 0, \overline{\partial})$ surrounding $\mathcal{O}_b$.  We can then take the OPE of the local and non-local operator.  In the case of sigma models that we are considering, the structure constant is the Schouten-Nijenhuis bracket on the corresponding polyvector fields.  This is an example of a secondary product (of the two local operators) described in \cite{Costello15, Beem:2018fng}.

%----------------------------------------------------------
\section{Product structure on cyclic homology}
%----------------------------------------------------------
At first it is surprising that the Schouten-Nijenhuis bracket corresponds to a particular certain subleading term in the OPE expansion.  However the BPS local operators that we consider have an interpretation as the operators in a holomorphic twist of the gauge theory.  Therefore Costello has suggested to consider the OPE directly in the holomorphically twisted theory.  We now illustrate this in the holomorphic twist of large-$N$ quiver gauge theories arising from branes at CY singularities following \cite{Eager:2018oww}.  For gauge theories arising from D-branes at singularities, a 
subset of operators correspond to the states contributing to the equivariant Hirzebruch $\chi_y$ genus of the infinity symmetric product of the vacuum

\begin{table}[htp]
\begin{center}
\begin{tabular}{|c|c|c|c|}
-1 & 0 & 1 & 2 \\
\hline
$\lambda$ & $\phi^n$ & $\psi_p$ & $f$
\end{tabular}
\end{center}
\caption{Homological grading of fields}
\label{tab:homologicalgrading}
\end{table}%

Consider the fields contributing to the 4d superconformal index of $\mathcal{N} = 4$ SYM as part of superfield
$$\Psi = \lambda + 2 \theta_n \phi^n + \epsilon^{mnp} \theta_m \theta_n \psi_p +  \theta_1 \theta_2 \theta_3 f $$
where the homological grading of the fields are shown in table \ref{tab:homologicalgrading}.
The homological degree zero piece of the potential term $\Psi \wedge \Psi \wedge \Psi$ of holomorphic Chern-Simons is
$$\Phi^{\flat} = \Phi + [\phi_n, \psi_n] \lambda + \lambda f^2$$
where $\Phi$ is the ordinary superpotential for the chiral fields $\phi_n$ in homological degree zero.    The graded potential $\Phi^{\flat}$ satisfies a non-commutative analog of the BV master equation
$$\{\Phi^{\flat}, \Phi^{\flat} \} = 0$$ 
where the bracket $\{ \cdot, \cdot \}$
$$\{ \Phi, \Psi \} = \sum_{n} \left( \pp{\Phi}{\phi_n}\pp{\Psi}{\psi_n} - \pp{\Phi}{\psi_n}\pp{\Psi}{\phi_n} \right)$$
is a noncommutative analog of the Schouten bracket.
We see that the fermions $\psi$ are in homological degree 1, and that the homological degree corresponds to the polyvector field degree.  The NC Schouten bracket then takes two elements of degree $p$ and $q$ to an element in degree $p + q - 1.$  The bracket corresponds to the product structure on cyclic homology.  In this case, the moduli space is the infinite symmetric product of a local CY manifold.
%----------------------------------------------------------
%----------------------------------------------------------
\section{Conclusions and outlook}
%----------------------------------------------------------
As promised, we have constructed local operators from polyvector fields on the quantum moduli space of vacua.  Our computation of the index in terms of these fields is somewhat miraculous that it appears to work at all.  Our computation the first step toward the full computation of the superconformal index of a 4d $\mathcal{N} = 1$ sigma model.  The full superconformal index can be expressed as an index over polyvector fields that arises from the holomorphic twist of the 4d $\mathcal{N} = 1$ sigma model.  The holomorphic twist of the sigma model is a four-dimensional $\beta\gamma$ system which is described in \cite{Aganagic:2017tvx}.  A natural further step is to complete the calculation of the superconformal index directly in terms of the four-dimensional $\beta\gamma$ system.  Perhaps there are cohomological vanishing theorems that can explain why our much simpler calculation comes so close to computing the exact result for the superconformal index.

We have considered the local operators in theories with moduli spaces of vacua that are cones over smooth K\"ahler bases.  A natural generalization is to consider more general cases where the moduli space has more complicated singularities or multiple branches.  In particular the Coulomb and Higgs branches of 4d $\mathcal{N} = 2$ and 3d $\mathcal{N} = 4$ gauge theories have been intensively studied.  While the Hilbert series of the Coulomb branch is captured by the Hall-Littlewood index \cite{Gadde:2011uv, Cremonesi:2013lqa, Razamat:2014pta} it is natural to wonder if more general indices can be recovered by considering polyvectors on the Coulomb and Higgs branches.  Recent work on the derived critical loci \cite{MR3728637} might help clarify the extra light states arising from singularities in the moduli space of vacua.

The super Lie algebra structure on polyvector fields is highly constraining and suggests a possible ``minibootstrap'' program for the short operators in 4d $\mathcal{N} = 1$ superconformal theories, similar to the 4d $\mathcal{N} = 2$ minibootstrap initiated in \cite{Beem:2014zpa}.  The super Lie algebra structure arising from the bracket structure on local operators is a higher dimensional analog of the bootstrap arguments given to determine the ground ring of two dimensional string theory \cite{Witten:1991zd} \cite{Lian:1992mn} which is a prototype for this type bootstrap program.

Similar to two-dimensional superconformal theories, it is natural to expect that when the moduli space of vacua has a global isometry group, the superconformal index can be expressed in terms of characters of a corresponding higher Kac-Moody group \cite{Kapranov:2017} \cite{Williams:2017}.  This extended symmetry could drastically simplify the ``minibootstrap'' program.

While we have focused on short BPS operators, an interesting series of papers \cite{Hellerman:2015nra}\cite{Hellerman:2017veg} determines the anomalous dimension of the lowest non-BPS operator with $R$-charge $J$ in a series expansion in $1/J$ directly from the effective theory on moduli space.  Another avenue to explore is the relationship with polyvector fields to possible WZW type interactions on the moduli space \cite{Manohar:1998iy, Intriligator:2000eq}.

While SQCD has a rich history and has been intensively studied for several decades, the new algebraic structures uncovered present progress towards solving a subset of the full theory.  Combined with new techniques from the bootstrap there is much hope to unravel more of the secrets of SQCD.

%-----------------------------------------------------------------
\section*{Acknowledgements}
%-----------------------------------------------------------------
The author thanks Christopher Beem, Kevin Costello, Sergey Galkin, Ingmar Saberi, Johannes Walcher, and Brian Williams for useful comments and discussions.  The computer algebra package \emph{Macaulay2} was used in the course of this work.
%-----------------------------------------------------------------
\appendix 
%----------------------------------------------------------

%----------------------------------------------------------
\section{Polyvectorfields from Borel-Weil-Bott}
\label{sec:BWB}
%----------------------------------------------------------
%----------------------------------------------------------
\paragraph{Borel-Weil-Bott for homogeneous bundles on Grassmannians}
%----------------------------------------------------------
The Borel-Weil-Bott Theorem computes the cohomology of line bundles on the flag variety of semisimple algebraic groups.  It can also be used to compute the cohomology of equivariant vector bundles on Grassmannians.

Let $V$ be a vector space of dimension $n$.  We identify the weight lattice of the group $GL(V)$ with $\Z^n.$
Let $X$ be the flag variety of $GL(V)$.  Let $L_{\lambda}$ denote the line bundle on $X$ corresponding to the weight $\lambda$.
Denote by
$$\rho = (n, n-1, \dots,2,1)$$
half the sum of the positive roots of $GL(V)$.  The corresponding line bundle
$L_{\rho}$ is the square root of the anticanonical line bundle.
\begin{thm}[Borel-Weil-Bott]
Assume that all entries of $\lambda + \rho$ are distinct.  Let $\sigma$ be the unique permutation such that $\sigma(\lambda + \rho)$ is strictly decreasing.  Then
$$H^k(X, L_{\lambda}) =
\begin{cases}
\Sigma^{\sigma(\lambda + \rho) - \rho} V^{*}, &  \text{if } k = \ell(\sigma) \\
0,&  \text{otherwise}\\
\end{cases}
$$
If not all entries of $\lambda + \rho$ are distinct then $\Hom^{\bullet}(X, L_{\lambda}) = 0$.
\end{thm}
%----------------------------------------------------------
\paragraph{Schur-Weyl modules}
%----------------------------------------------------------
We fix a vector space $V$ of dimension $n.$  Let $\lambda = (\lambda_1, \dots, \lambda_s)$ be a partition of $n$.
Then the Schur-Weyl module $\mathbb{S}_{\lambda} V$ is an irreducible representation of $GL(V)$.
It can be explicitly described as
$$\mathbb{S}_{\lambda} V = \frac{\bigwedge^{\lambda_1} V \otimes \bigwedge^{\lambda_2} V \otimes \dots \otimes \bigwedge^{\lambda_s} V}{R(\lambda, V)}$$
where $R(\lambda, V)$ is a subvector space defined in section 2.1 of \cite{MR1988690}.
For two vector spaces $S$ and $Q,$ there is an isomorphism of $GL(S) \times GL(Q)$  equivariant representations
$$\bigwedge^r (S \otimes Q) \cong \bigoplus_{|\lambda| = r} \mathbb{S}_{\lambda} S \otimes \mathbb{S}_{\lambda'} Q$$
where $\lambda'$ is the conjugate partition of $\lambda$ which is corollary (2.3.3) of \cite{MR1988690}.
Recall that $T_{\Gr} \cong \Hom_{\Gr}(S, Q) \cong S^{*} \otimes Q,$ where $S^{*}$ is the dual of the subbundle.
$$\bigwedge^r T_{\Gr} \cong \bigoplus_{|\lambda| = r} \mathbb{S}_{\lambda} S^{*} \otimes \mathbb{S}_{\lambda'} Q$$
We can compute the cohomology of the bundles $\wedge^r T_{\Gr}$ using the following theorem.
\begin{thm}[Borel-Weil-Bott for Grassmannians]
Let $\Gr(k, V)$ be the Grassmannian of $k$-planes in a $n$-dimensional vector space $V.$
Consider the vector bundle $E_{\alpha,\beta} =  \mathbb{S}_{\alpha} Q \otimes \mathbb{S}_{\beta} S$ on $\Gr(k, V)$
where $\alpha = (\alpha_1, \dots, \alpha_{n-k})$ and $\beta = (\beta_1, \dots, \beta_{k})$ are partitions of length at most $n - k$ and $k$ respectively.
Let $\lambda = (\alpha, \beta)$ and $\rho = (n, n-1, \dots,2,1)$ be half the sum of the positive roots of $GL(V)$.   There is a unique non-vanishing cohomology group
\begin{equation}
H^{k}(\Gr, E_{\alpha,\beta} )\cong 
\begin{cases}
\mathbb{S}_{\sigma(\lambda+ \rho) - \rho} V &  \text{if } k = \ell(\sigma)\\
0,&  \text{otherwise}\\
\end{cases}
\end{equation}
where $\sigma$ is the unique element of the Weyl group that take $\lambda+ \rho$ to a dominant weight
and $ \ell(\sigma)$ is the number of adjacent transpositions (simple Weyl reflections) needed to form $\sigma.$
Furthermore, the isomorphism is $SL(V)$-equivariant.
\end{thm}  For a proof see \cite{MR1988690} Corollary 4.1.9.
\newpage
\section{Tables of polyvector fields}
\label{sec:PVF}
%----------------------------------------------------------

\begin{table}[h]
\tiny
\begin{center}
\begin{tabular}{c||c|c|c|c|c|c|c|c|c||c|}
dim & $\O_X$ & $TX$ & $\wedge^2 TX$ & $\wedge^3 TX$  & $\wedge^4 TX$ & $\wedge^5 TX$ & $\wedge^6 TX$ & $\wedge^7 TX$ & $\wedge^8 TX$ & \\
\hline
0 & 1 & 0 & 0 & 0 & 0 & 0 & 0 & 0 & 0 & 1\\
\hline
1 & 15 & 0 & 0 & 0 & 0 & 0 & 0 & 0 & 0 & 15 \\
\hline
2 & 105 & 0 & 0 & 0 & 0 & 0 & 0 & 0 & 0 & 105 \\
\hline
3 & 490 & -35 & 0 & 0 & 0 & 0 & 0 & 0 & 0 & 455 \\
\hline
4 & 1764 & -384 & 0 & 0 & 0 & 0 & 0 & 0 & 0 & 1380 \\
\hline
5 & 5292 & -2205 & 21 & 0 & 0 & 0 & 0 & 0 & 0 & 3108 \\
\hline
6 & 13860 & -8960 & 560 & 0 & 0 & 0 & 0 & 0 & 0 & 5460 \\
\hline
7 & 32670 & -29106 & 4230 & 21 & 0 & 0 & 0 & 0 & 0 & 7815 \\
\hline
8 & 70785 & -80640 & 19782 & -384 & 0 & 0 & 0 & 0 & 0 & 9543 \\
\hline
9 & 143143 & -198198 & 70070 & -4515 & -35 & 0 & 0 & 0 & 0 & 10465\\
\hline
10 & 273273 & -443520 & 205920 & -24960 & 105 & 0 & 0 & 0 & 0 & 10818 \\
\hline
11 & 496860 & -920205 & 528255 & -97020 &  3003 & 15 & 0 & 0 & 0 & 10908 \\
\hline
12 & 866320 & -1793792 & 1221220 & -302848 &  20020 & 0 & 0 & 0 & 0 & 10920 \\
\hline
13 & 1456560 & -3318315 & 2599443 & -810810 & 85410&  -1365 & 0 & 0 & 0 & 10923 \\
\hline
14 & 2372112 & -5870592 & 5172960 & -1935360 &  282555 & -10752 & 0 & 0 & 0 & 10923 \\
\end{tabular}
\end{center}
\caption{$\chi$ of polyvector fields on $Gr(2,6)$.}
\label{pvfieldsgr26}
\end{table}%

\begin{table}[h]
\tiny
\begin{center}
\begin{tabular}{c|c|c|c|c|c|c|c|c|c|c|c|c|}
dim & $\O_X$ & $TX$ & $\wedge^2 TX$ & $\wedge^3 TX$  & $\wedge^4 TX$ & $\wedge^5 TX$ & $\wedge^6 TX$ & $\wedge^7 TX$ & $\wedge^8 TX$ & $\wedge^8 TX$ & $\wedge^8 TX$  & \\
0 & 1 & 0 & 0 & 0 & 0 & 0 & 0 & 0 & 0 & 0 & 0 &  1 \\ 
1 & 28 & 0 & 0 & 0 & 0 & 0 & 0 & 0 & 0 & 0 & 0 &  28 \\
2 & 336 & - 63 & 0 & 0 & 0 & 0 & 0 & 0 & 0 & 0 & 0 & 273 \\
 3 & 2520 & -1280 & 36 & 0 & 0 & 0 & 0 & 0 & 0 & 0 & 0 &  1276 \\ 
4 & 13860 & -12474 & 1890 & 0 & 0 & 0 & 0 & 0 & 0 & 0 & 0 &  3276 \\ 
5 & 60984 & - 80640 & 26180 & - 1280 & 36 & 0 & 0 & 0 & 0 & 0 & 0 &  5280 \\
6 & 226512 & - 396396 & 205920 & - 30030 & 336 & - 63 & 0 & 0 & 0 & 0 & 0 & 6279 \\ 
7 & 736164 & - 1596672 & 1149876 & - 302848 & 19980 & 0 & 28 & 0 & 0 & 0 & 0 &  6528 \\ 
8 & 2147145 & - 5521230 & 5069064 & - 1963962 & 282555 & - 7020 & 0 & 0 & 0 & 0 & 0 &  6552 \\ 
9 & 5725720 & - 16912896 & 18752580 & - 9580800 & 2194500 & - 172800 & 252 & 0 & 0 & 0 & 0 &  6556 \\
10 & 14158144 & - 46930455 & 60540480 & - 38153115 & 12008304 & - 1686069 & 68320 & 945 & 0 & 0 & 0 &  6554 \\
11 & 32821152 & - 119927808 & 175207032 & - 130296320 & 51832872 & - 10526208 & 912384 & - 16128 & - 420 & 0 & 0 & 6556 \\ 
12 & 71954064 & - 285817532 & 463440978 & - 394215822 & 187979220 & - 49621572 & 6637428 & - 352044 & 1764 & 70 & 0 & 6554 \\ 
\end{tabular}
\end{center}
\caption{$\chi$ of polyvector fields on $Gr(2,8)$.}
\label{pvfieldsGr28C}
\end{table}%

\cleardoublepage %force tables to print before the bibliography
\bibliographystyle{ytphys}
\bibliography{HoloTwist}
%----------------------------------------------------------
\end{document}